\let\texyear\year

\documentclass{ieeeaccess}
\usepackage{cite}
\usepackage{amsmath,amssymb,amsfonts}
\usepackage{algorithmic}
\usepackage{graphicx}
\usepackage{textcomp}
\usepackage{multirow}

\usepackage{hyperref}
\hypersetup{
    colorlinks=true,
    linkcolor=blue,
    filecolor=magenta,      
    urlcolor=cyan,
    pdftitle={Overleaf Example},
    pdfpagemode=FullScreen,
    }
    
\usepackage[ruled,vlined]{algorithm2e}
\let\ieeeaccessyear\year
\let\year\texyear
\usepackage{forest}
\usepackage{pgfplots}
\usepackage{tcolorbox}
\let\year\ieeeaccessyear
\definecolor{accessblue}{RGB}{0,105,154}

\def\BibTeX{{\rm B\kern-.05em{\sc i\kern-.025em b}\kern-.08em
    T\kern-.1667em\lower.7ex\hbox{E}\kern-.125emX}}

\newcommand{\tool}[1]{\textsc{#1}\xspace}

\newcommand{\sherlock}{\tool{SHERLOCK}}

\newcommand{\accuracybinary}{97\%}
\newcommand{\macrofamily}{.491}
\newcommand{\macrotype}{.497}

\pgfplotsset{compat=1.11,
        /pgfplots/ybar legend/.style={
        /pgfplots/legend image code/.code={%
        \draw[##1,/tikz/.cd,bar width=3pt,yshift=-0.2em,bar shift=0pt]
                plot coordinates {(0cm,0.8em)};},
},
}

\pgfplotsset{compat=1.11,
        /pgfplots/xbar legend/.style={
        /pgfplots/legend image code/.code={%
        \draw[##1,/tikz/.cd,bar width=5pt,yshift=-0.4em,bar shift=0pt]
                plot coordinates {(0.4cm,0.4cm)};},
},
}

\begin{document}

\history{Date of publication xxxx 00, 0000, date of current version xxxx 00, 0000.}

\doi{10.1109/ACCESS.2017.DOI}

\title{Self-Supervised Vision Transformers for Malware Detection}

\author{\uppercase{Sachith Seneviratne}\authorrefmark{1},
\uppercase{Ridwan Shariffdeen\authorrefmark{2}, Sanka Rasnayaka\authorrefmark{2}, and}
\uppercase{Nuran Kasthuriarachchi\authorrefmark{3}}.}
\address[1]{University Of Melbourne, Australia (e-mail: sachith.seneviratne@unimelb.edu.au)}
\address[2]{School of Computing, National University of Singapore (e-mail: ridwan@u.nus.edu, sanka@nus.edu.sg)}
\address[3]{University of Moratuwa, Sri Lanka (e-mail: nuran.11@cse.mrt.ac.lk)}
\tfootnote{This work was supported by Australian NHMRC Grant GA80134.}

\markboth
{Author \headeretal: Preparation of Papers for IEEE TRANSACTIONS and JOURNALS}
{Author \headeretal: Preparation of Papers for IEEE TRANSACTIONS and JOURNALS}

\corresp{Corresponding author: Sachith Seneviratne (e-mail: sachith.seneviratne@unimelb.edu.au).}

\begin{abstract}
Malware detection plays a crucial role in cyber-security with the increase in malware growth and advancements in cyber-attacks. Previously unseen malware which is not determined by security vendors are often used in these attacks and it is becoming inevitable to find a solution that can self-learn from unlabeled sample data. This paper presents \sherlock, a self-supervision based deep learning model to detect malware based on the Vision Transformer (ViT) architecture. \sherlock ~is a novel malware detection method which learns unique features to differentiate malware from benign programs with the use of image-based binary representation. Experimental results using 1.2 million Android applications across a hierarchy of 47 types and 696 families, shows that self-supervised learning can achieve an accuracy of \accuracybinary~ for the binary classification of malware which is higher than existing state-of-the-art techniques. Our proposed model is also able to outperform state-of-the-art techniques for multi-class malware classification of types and family with macro-F1 score of \macrotype~ and \macrofamily~ respectively. 
\end{abstract}


\titlepgskip=-15pt

\maketitle

\section{Introduction}
\label{sec:introduction}

\newlength{\xfigwd}
\setlength{\xfigwd}{\linewidth}
\PARstart{A}{rtificial} intelligence has gained significant popularity over the recent years, serving in many applications since its inception. Currently, it dominates in the imaging field - in particular, image classification~\cite{al2017review}. Advancements in deep learning technology has further improved the accuracy and reliability of models that use deep learning architectures. One can be impressed by the performance of Deep Neural Networks (DNNs) which takes advantage of the proliferation of large data-sets in addition to the increase in computational power. Despite state-of-the-art performance achieved by DNNs, recent research reveals that such models are prone to adversarial attacks which alters the output of the DNN and degrades the performance~\cite{subramanya2019fooling}. Adversarial perturbation is a well-known problem, due to the nature of deep neural networks which are highly sensitive to slight modifications that are imperceptible to human eye. Recently, many works have been proposed to generate diverse set of adversarial examples, which have been used as benchmarks to evaluate the robustness of deep learning models~\cite{dong2020benchmarking}. Similarly, several strategies are also bought forward in the literature to thwart such attacks~\cite{qiu2019review}. 

In the realm of software security, malware detection poses an increasingly challenging task~\cite{zhu2016featuresmith}, which is to identify malicious software programs by differentiating such programs apart from benign applications. According to recent studies~\cite{lee2022study}, malicious software is increasing at an alarming rate, where some or most hide inside popular applications using obfuscation techniques that evades traditional signature-based detection methods. In particular, Android malware detection plays a vital role which seriously threatens the integrity of Android applications. Statistics shows that more than 3.25 million malicious Android apps have been found in 2016~\cite{omer2021efficiency}, indicating that a new malware app is found every 10 seconds. Android malware detection needs more attention and research to prevent malware from being released in the wild through Android application marketplaces. Malware authors use common obfuscation techniques such as encryption, oligomorphic, polymorphic, metamorphic, stealth and packaging methods to make the task of detection difficult~\cite{deng2022polymorphic}. The malware detection problem can be seen as a classification problem whereby the algorithm needs to correctly categorize a given program into one of the classes: malware or benign. Most existing approaches to detect Android malware rely on extraction of application features and using prior knowledge to train a machine learning classifier to distinguish between benign and malicious applications. Despite the high accuracy, these models are ineffective as attackers bypass such detection by adding features commonly used by benign apps. Such simple addition of benign functionality to malicious apps, such as pop-up messages and logging, can change the detection class from malicious to benign. 

Recent malware programs are written with the objective of being highly similar to benign applications by incorporating features of popular applications, having only slight differences which is difficult for humans to perceive~\cite{faruki2014android}. DNNs can detect such imperceptible changes, and by training with a large data-set we can detect features of malicious applications despite having a large amount of benign features embedded. In fact, previous research has shown that malware can be detected using image classifiers made of deep learning models~\cite{Rezende2017}.  Malware variants belonging to the same family exhibits visual similarity in the byteplot images, which can be used to train deep learning models to detect specific features which can then be used to detect the existence of similar variant in other applications. These images are generated based on visualization in the spatial domain by converting bytes to pixels. The works typically convert malware binaries to digital images and pass them into a neural network in order to detect malware. Machine learning is widely applied in the detection of Android malware, whether based on static, dynamic or hybrid analysis approaches. Compared with traditional methods, such as signature-based detection, machine-learning based approaches provide better performance in detection efficacy and efficiency, in addition to the ability to detect previously unseen types of malware i.e. zero-shot learning. 

In this paper, we present a novel technique building on self-supervised representation learning to detect malware with high-accuracy surpassing state-of-the-art techniques on one of the largest malware data-sets containing 1.2 million binary images across a hierarchy of 47 types and 696 families of malware. Self-supervised learning is capable of adopting self-defined pseudo-labels as supervision and use the learned representations to avoid the cost of annotating large-scale data-sets. Traditional supervised learning methods heavily depend on the availability of annotated large-scale training data, which is impractical with the ever-evolving nature of malware programs. Even though there are plethora of data available in a limited scale, the learning can suffer from issues such as generalization error, spurious correlations and adversarial attacks. This is where self-supervised methods play a vital role in reaping the benefits of deep learning without the expense of heavy annotations and learn feature representations where data provides the supervision. We show that our proposed technique is able to achieve 97\% accuracy in detecting malware and with 87\% precision in correctly identifying the malware family, outperforming state-of-the-art techniques for one of the largest data-set consisting of million Android apps curated from AndroZoo~\cite{androzoo} repository. The main contributions of this study are as follows:

\begin{itemize}
    \item Implementation of a Transformer based computer vision model utilizing self-supervised learning for Android malware detection. To the best of our knowledge this is the first work to utilize self-supervised learning and vision transformers for the purpose of detecting malware on a large-scale data-set.
    \item A comprehensive evaluation of the effectiveness in detecting a malware and multi-class malware classification into malware type and malware family. Our results demonstrate that self-supervised learning can effectively classify 47 types and 696 families of malware with a macro-F1 score of 0.497 and 0.491 respectively.  
    \item Contributing to the understanding of self-supervised computer vision models, specifically the Vision Transformer architectures, and their performance on synthetic imagery.
\end{itemize}

The rest of the article is organized as follows: Section~\ref{sec:background} provides a background to malware detection and self-supervised learning. Next, Section~\ref{sec:methodology} details the methodology while Section~\ref{sec:evaluation} elaborates on the evaluation and the results. Section~\ref{sec:related} gives an overview of related work with Section~\ref{sec:perspectives} discussing important insights. Finally, the conclusion is given at the end of the article. 
\section{Background}
\label{sec:background}
In this section, we summarize the important background details for self-supervised malware detection using byteplot grayscale images. First, we describe the process for the image generation where the malware executable is converted to a byteplot grayscale image. These converted images of binary executables can be used to automatically identify visual patterns that can be used in static malware analysis. We then describe the necessary background on image classification using machine learning and self-supervision for image classification. 

\subsection{Byteplot Visualization}
Transforming a binary executable into an image was first described by Conti et al~\cite{conti2008visual} to represent binary data objects as grayscale images, where a pixel value of the image corresponds to a byte in the binary. Since a byte range is 0-255, the image pixel color is rendered as a grayscale where zero is black and 255 is white other values representing intermediate shades of gray. This representation provides a visual analysis of binary data to distinguish structurally different regions of data and thus enabling wide range of static analysis such as file type identification, inference of primitive data type, fragment classification and other tasks that required special reverse engineering effort for the binary data. Nataraj et al proposed to transform malware samples into byteplot images with the observation that significant visual similarities exists in malware belonging to the same family~\cite{nataraj11}. The transformation consist of reading the binary as a vector of 8-bit unsigned integers organized into a 2D array, where the width is defined based on empirical observations and height is varied depending on the file size. 

Freitas et al extended this work for Android applications by constructing the image representation using DEX file (bytecode) obtained from the Android APK~\cite{malnetimages}. The extracted DEX file was converted to a 1D array (instead of 2D) of 8-bit unsigned integers. The vector is then used in a 3-phase conversion process of 1) converting the 1D array to a 2D image representation 2) scaling the image to a standard size and 3) encoding semantic information into the RGB channels. Similar to the approach presented by Nataraj et al~\cite{nataraj11} the first step is to convert the data array into a grayscale image, where the width is fixed and height is varied based on the file size. The transformed image is then scaled to $256\times256$ using a standard Lanczos filter from the Pillow library. Once the scaled grayscale image is obtained the final step is encoding semantic information that can further assist the differentiation of distinctive texture patterns. By coloring each byte according to its usage in the malware executable file, the image has an added layer of semantic information on top of the raw bytecode. Gamut et al present~\cite{gennissen2017gamut} an encoding that assigns each byte to a particular RGB color channel depending on its position in the DEX file structure. The observation is that the sections in a DEX file should be distinguished for malware detection as each malware type utilize each section differently, and such patterns can be easily recognized in an image visualization. The encoded image can be decoded back to grayscale by combining each of the channels. 

\subsection{Image Classification}
Image classification is a paradigm of machine learning involving assigning categories to images. Most image classification problems are multi-class, single label classification problems involving multiple potential categories for the images but with only a single label or category being applied to an image. However, it is possible to apply multiple labels to some imagery. For example, images that contain multiple objects (i.e. a vehicle and a horse) or images that captures multiple features (i.e. an image of a bird can be labeled based on its physical features). In the context of malware analysis, a byteplot image generated from a malware program may contain different labels such as, if its benign or malware, type of malware, author of the program and malware family. These labels can be explored as categories either independently (as multiple single-label, multi-class problems) or jointly (a single multi-label, multi-class problem). A key issue when performing multiple single-label, multi-class classifications on the same dataset is the computational cost associated with training a different network for each problem. Transfer learning is a commonly used technique to alleviate some of the computational cost. In transfer learning, a previously trained neural network is reused to instantiate the weights of a new neural network that is then trained on the new problem. By reusing the weights of the previous task the "knowledge" from the previous task is transferred over and reused to some extent. Rezende et al~\cite{Rezende2017} used transfer learning to train a DNN using ImageNet dataset which contains 1.28 million images of 1000 classes (which does not include malware images) and was able to achieve an accuracy of 92.97\% on a dataset comprising 10,136 malware byteplot images. This approach was one of the first to show that, without feature engineering, using raw pixel values of byteplot images, malware can be classified with high accuracy. The DNN was initially trained for a different task on a different dataset, yet was capable to outperform at the time state-of-the-art models that used specialized reverse engineering efforts. This line of work depend on the learning from already established, annotated data-sets despite the disparity between the initial training data-set used for the weights and the malware data-set used for testing. However, Freitas et al~\cite{malnetimages} shows that transfer learning does not scale when there is a significant disparity between the training data-set and testing data-set. Using a ResNet18 model pre-trained on ImageNet and fine tuning for malware images, they show that the pre-trained model on ImageNet is less effective than a model trained from scratch on a malware data. In order to achieve the full potential of image classification for malware analysis, the model must learn features from a data-set consisting of malware images. 

The challenge is to generate a large data-set for malware images with annotations on different labels it can learn. Generating such a large-scale data-set may not be feasible as it was for ImageNet because the manual annotation for images in ImageNet can be often done by individuals with no special expertise and with minimal time. However, the classification of a malware requires an expert analyst on average around 10 hours to characterize a malicious binary~\cite{mohaisen2014chatter, votipka2020observational}. The expensive effort makes it impractical to generate a large corpus of malware images that consist of expert-labeled data that can be used for different tasks such as classification for malware family, type of malware, author of malware etc. Given the evolution of software with new API changes, new code generation techniques and improved compiler optimizations, program binaries as a whole changes over time. Similarly, adversaries design malicious binaries to be adaptable to these changes while also avoiding detection~\cite{kantchelian2015better}. This makes it challenging to maintain an up-to-date data-set consisting of expert-labeled data, that can be used to train to detect new malware with new features. This is a concept known a ``conceptual drift'' where an expert-labeled data-set for malware detection can quickly be outdated due to the nature of malware evolution.

\begin{figure*}[t]
\centering
\includegraphics[width=0.8\textwidth]{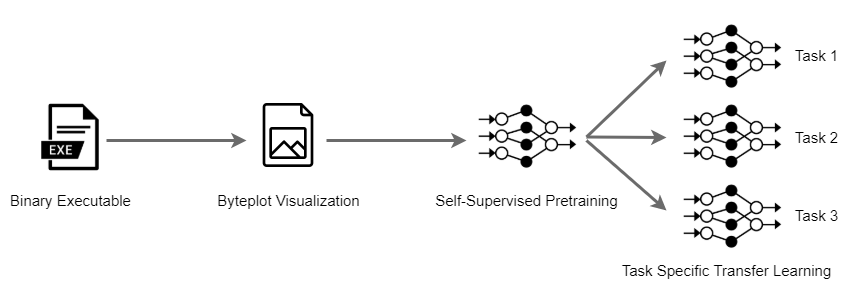}
\caption{A high-level overview of \sherlock: self-supervised learning can be used to bootstrap different classification problems using malware image data}
\label{fig:overview}
\end{figure*}

\subsection{Self-Supervised Learning}
When labeled data is scarce, self-supervised learning holds great promise for improving representation learning. Self supervision is a paradigm of representation learning which involves training a model on a pretext task with unlabeled imagery, but in a supervised manner. The pretext tasks in this context are a class of tasks which can be automatically generated from imagery without human intervention. The goal of self-supervised learning from images is to construct image representations that have semantic meaning via pretext tasks that does not require any manual annotations. Examples of pretext tasks include jigsaw puzzle solving\cite{noroozi2016unsupervised} for reconstructing an image after randomly permuting grid sections, relative position prediction\cite{doersch2015unsupervised} for predicting the relative position and orientation of image patches with respect to each other or image colorization\cite{zhang2016colorful} for removing colour from an image and restoring it. These tasks require the neural network to learn a generic representation of the dataset in order to solve the pretext task. A model trained for solving these pretext tasks learn representations that can be reused for solving other downstream tasks of interest, such as image classification. 

Purely self-supervised techniques learn visual representations that may be inferior to those achieved via fully-supervised techniques on a given downstream task. Thus, self-supervision alone is insufficient for practical use despite demonstrations of encouraging results in recent work. To alleviate this, self-supervision is combined with transfer learning or fine-tuning on the downstream task. This is similar to how humans perform learning. We bring an innate representation of the world to any new task we learn. In the self-supervised learning setting, the labelled imagery in the downstream task provides a supervisory signal (usually by propogating an error function such as cross entropy) which helps the self-supervised representation to adapt to the downstream task. Zhai et al ~\cite{zhai2019s4l} shows that self-supervised learning can dramatically benefit from a small amount of labeled examples by turning self-supervised to semi-supervised learning model. Such a self-supervised learning can alleviate the need for continuously generating manually annotated dataset for malware classification. Moreover, self-supervision can improve several aspects of model robustness, including robustness to adversarial examples, label corruptions and common input corruptions. Hendrycks et al ~\cite{hendrycks2019using} shows that self-supervised learning can even surpass fully supervised methods, raising the question of whether self-supervised learning render large labeled data-sets needless. However, in this work, we investigate the benefits of self-supervised learning combined with a large-scaled labeled data-set, specifically for malware detection and classification. In particular, we explore how label-independent learning can provide efficient workflows for performing multiple single label classifications on the same dataset.

While self-supervised computer vision has received considerable attention, most current work is focused on natural imagery (imagery captured from a camera of a natural scene). However, many forms of imagery consumed by humans take on a more abstract form. There are many concepts that are visually represented as images. The underlying dynamics of such imagery can be drastically different from that of natural scene imagery. For example, an image of a natural scene contains semantic cues as to the proximity of an object to the sensor based on the size of the subject. However, such semantics are noticeably absent or different in more abstract imagery. Additionally the transferability of knowledge representations would be further lowered between abstract imagery domains due to this reason, when compared with natural imagery domains, which points to an increased utility in developing approaches such as self-supervised learning for such domains. Further understanding of how self-supervised learning operates on such imagery is therefore of considerable academic interest. Some prior work in this regard\cite{xie2019unsupervised,bhunia2021vectorization} focuses on sketch based abstractions of images, while others operate on images of maps\cite{seneviratne2021self} or digital elevation outputs\cite{seneviratne2021contrastive}. To the best of our knowledge, no prior work has performed self-supervised learning on images representing binaries of programs which falls into this category of imagery as it is an abstract representation of the underlying software artifact.

\section{Methodology}
\label{sec:methodology}
In this section we describe the overall architecture of our model and an overview of our proposed method to detect and classify malware using self-supervised learning. Next, we describe self-supervised learning to synthesise malware images using a new architecture in computer vision. Finally, we detail how we fine-tune the model by reusing the learnt representation from the synthesis task, for the problem of multi-class labelling in the context of malware classification.

\subsection{Overview}
We propose a novel technique for malware detection using the vision transformer architecture with self-supervised learning. Compared to existing techniques that use image classification for malware detection, we interpret a malware image as a sequence of patches and process it by a standard Transformer encoder as used in natural language processing. This interpretation combined with pre-training on large data-sets allows us to re-purpose a trained model effectively for 3 different classification tasks with minimal computation cost. Figure~\ref{fig:overview} depicts an overview of our proposed method, \sherlock. The core idea of \sherlock~ is to generate training samples for the underlying learning task in a fully-supervised manner, which then can be used by classification models to bootstrap the training via transfer learning. The learning task is to synthesise malware images from a limited set of features that captures the semantics of a malware. This is contrary to the previous work on malware detection which is entirely based on supervised learning.

\sherlock~consists of two main components that can efficiently synthesise malware images using an optimized representation of the malware image and a classification component that can reuse the optimized representation to identify the correct label for a specific classification problem. Given a malware image, \sherlock~first splits the image into patches (small pieces of the image) of size $16\times16$ pixels. These patches are then flattened or concatenated to form a 1-dimensional vector, which is used to generate a lower dimensional linear embedding representation of the patches. As transformers are used for modelling sequences, positional embedding is used to maintain the 2-dimensional positional correspondence of the image patches. Using such an embedding we train a model to encode the semantic features of a malware image in a fully-supervised manner to obtain an optimized version of the embedding. This optimized embedding is used to bootstrap the classifier component as a base knowledge that captures the semantic encoding of a malware, and fine-tune different classification models on malware image data (i.e. malware detection, type-classification, family-classification).
We use the MalNet dataset\cite{malnetimages} in our analysis as it is the largest publicly available cybersecurity image database (1.2 million images) with multiple labels for each image corresponding to malware/benign (2 categories), malware type (47 categories) and malware family (696 categories).

\subsection{Masked Auto Encoder}
\label{sec:MAE_selfsup}
\begin{figure}[t]
\centering
\includegraphics[width=0.10\textwidth]{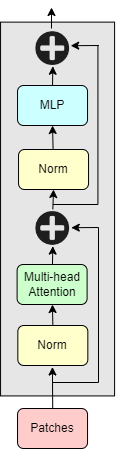}
\caption{Illustration of the Transformer Encoder \cite{dosovitskiy2020image} that takes a sequence of patches as input and transforms into an encoding.}
\label{transformerencoder}
\end{figure}

The recent advent of Vision Transformers\cite{dosovitskiy2020image} (ViTs) provide an alternate architecture for image processing that is both simple and scalable. In a Vision Transformer, an image is first split into patches (of sizes such as $8\times8$ or $16\times16$ pixels). These patches are then transformed into an embedding using standard transformer encoders as shown in Figure~\ref{transformerencoder}. Given an image, the model encodes a representation which it then decodes to regenerate the original image. Therefore, the input to the model is an image, and the output is a reconstruction of the same image, from a reduced feature space with a perfectly accurate output being an exact copy of the original image. Encoding the image into a bottleneck representation (which is usually a fixed size vector of size 256, 512, 1048 etc.) enforces the self-supervisory task  to learn useful semantic features for the decoder which is used to reconstruct the image. This concept can be generalized to other modalities of data, for example, Natural Language Processing (NLP) where masked language modelling as a self-supervisory task has previously led to state of the art results~\cite{devlin-etal-2019-bert}.

\begin{figure}[t]
\centering
\resizebox{0.48\textwidth}{!}{%
\begin{tabular}{ccccc}
Type & Original & Mask & Masked & Synthesized \\
trojan/triada &
\includegraphics[width = 1in]{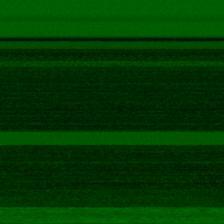} &
\includegraphics[width = 1in]{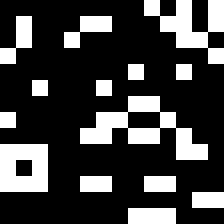} &
\includegraphics[width = 1in]{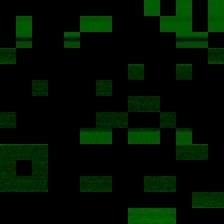} &
\includegraphics[width = 1in]{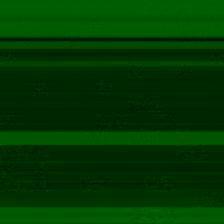}\\

addisplay/adflex &
\includegraphics[width = 1in]{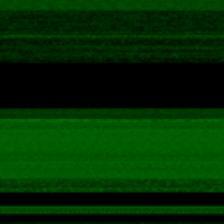} &
\includegraphics[width = 1in]{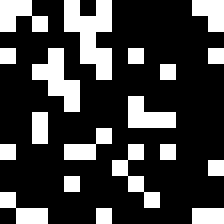} &
\includegraphics[width = 1in]{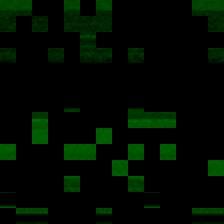} &
\includegraphics[width = 1in]{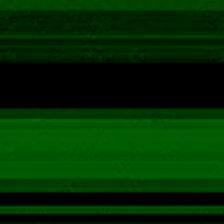}\\

\end{tabular}

}
\caption{Illustration of the image synthesis using images of two malware types. Using an masked image which masks 75\% of the image our learning model is able to synthesise images almost identical to the original image.}
\label{fig:image-synthesis}
\end{figure}

We utilize masked auto encoding as our pre-text task. In general masked auto encoding is associated with removing a portion of the data, and using the remaining data to predict the removed portion. Thus, given an image, some pixels of the image are masked prior to self-supervised training and the neural network must learn how to recover the masked pixels based on the visible pixels. Recent work\cite{he2021masked} has shown that masking a large portion of pixels in natural images (such as ImageNet) leads to a challenging self-supervisory task capable of generating useful representations for downstream tasks. In the absence of prior work applying masked auto-encoding to abstract/synthetic imagery, we explore a large masked pixel percentage (75\% of the image), thus forcing the model to attempt to recover the image based only on the unmasked 25\% of the image, such high percentages have been shown to work well for natural imagery \cite{he2021masked}. Figure~\ref{fig:image-synthesis} depicts the results of the image synthesis for two different malware types. Importantly, we use Vision Transformers \cite{dosovitskiy2020image} in our analysis, which are able to effectively incorporate 2-D positional embeddings in order to indicate the location of a patch within an image. Such incorporation is not straightforward in convolutional neural networks. 


\begin{figure*}[t]
\centering
\includegraphics[width=\textwidth]{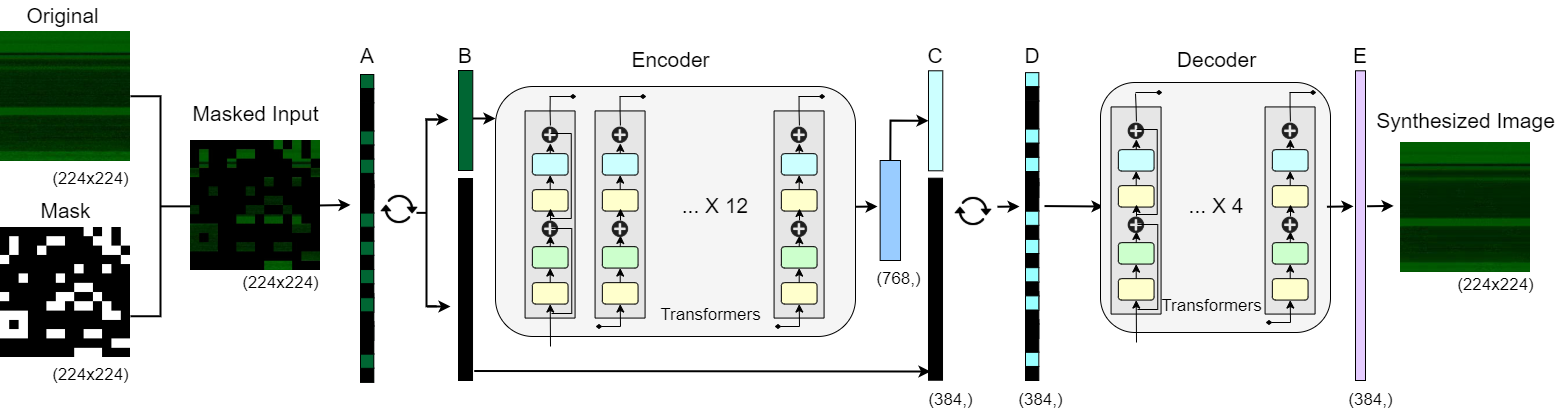}
\caption{Overview of the masked auto encoder which takes masked images as input and learns to reconstruct the original image using self-supervised learning. Layer A is the linearization of the patches which is then shuffled in Layer B to separate masked and unmasked pixels. Layer C depicts the embedding of the masked and unmasked pixels. Layer D represents combined encoding which is the input layer for the decoder to learn. Layer E is the final embedding for the reconstruction of the original image. }
\label{fig:maskedautoencode}
\end{figure*}
Figure~\ref{fig:maskedautoencode} gives an overview of the auto-encoder component of \sherlock~that use self-supervised learning to obtain an optimized embedding to reconstruct masked images with a higher accuracy. A mask is generated by uniformly sampling 25\% of the image to retain. The original image is then filtered using the generated mask to generate the masked input image. The masked input image is flattened to generate embedding for each patch. Each image is $224\times224$ with 3 RGB channels, each patch is $16\times16$ leading to $14\times14 = 196$ patches in total. Each embedding is a vector of size $3\times16\times16 = 768$. The embeddings are shuffled to separate the masked vs unmasked patches, with only the unmasked patches being passed through the encoder architecture. The unmasked embeddings are passed through 12 architecturally identical blocks which have different weights in order to generate an embedding of size 768 for each patch. The unmasked encoder embeddings (size 768) are converted into unmasked decoder embeddings (size 384) using a linear projection layer to match the width of the decoder. The unmasked decoder embeddings are combined with the flattened masked patches, which are represented by a common, learnable mask token indicating the presence of a masked patch. This indicates to the model that such patches are to be predicted in the decoding process. The embeddings and mask tokens are decoded by the decoder, which contains 4 architecturally identical blocks with different weights. In order to generate a final embedding for each patch (size 384). This embedding is passed through a linear projection which maps it to the number of output pixels in a patch ($3\times16\times16 = 768$. At this stage, a pixel-wise loss is computed between the original image and the synthesized image, which is then backpropagated throughout the network.

Importantly, the ability to only encode the unmasked inputs enables more computationally efficient training in a self-supervised manner. The decoder is only used during self-supervised pre-training which allows flexibility in decoder design. In particular, simpler decoders enforce a better representation in the other parts of the architecture as less knowledge/parameters would be stored within the decoder. The encoder containers 12 blocks while the decoder only has 4, additionally the embedding dimension for the encoder has size 768 while the decoder has size 384. To evaluate reconstruction error, each pixel in a patch is first normalized by subtracting the mean of pixels in the patch and dividing the result by the patch standard deviation. The reconstruction loss is computed by calculating mean standard error over pixels in the masked patches after normalization.


\subsection{Classification Models}
One of the major advantage in utilizing a self-supervised learning in this manner is the possibility to reuse the same self-supervised representation with minimal additional training in all 3 downstream tasks - in this case malware classification (binary), malware type classification (47 categories) and malware family classification (696 categories). In contrast, simple supervised models would need to separately train from scratch, which would require considerable additional computational processing. In this regard, as more tasks that are formulated on the same dataset, self-supervised methods become more efficient. The self-supervised representations generated are thus fine-tuned separately on each individual task with accuracy and macro f1 scores generated for each task.

\begin{figure}[h]
\centering
\includegraphics[width=0.5\textwidth]{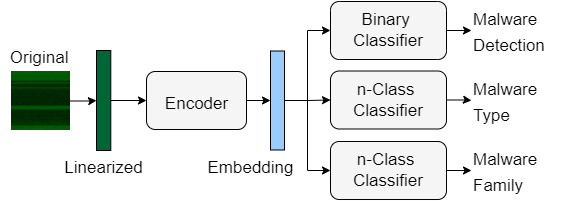}
\caption{Overview of the classification models in \sherlock, which reuse the optimized embedding generated via self-supervised learning to bootstrap 3 different classification tasks. }
\label{fig:classifier}
\end{figure}

\begin{algorithm}
\label{Alg:sherlock}
 \caption{Malware Detection algorithm}
\SetAlgoLined
\textbf{Input}: Binary Image, $I_{raw}$\\
\textbf{Output}: Boolean value true/false  \\
\textbf{Execution}:\\
  1. $I$ = $resize(I_{raw}$)\\ 
  2. $L_{patches}$ = $(I)$\\
  3. $encoding$ = $encode(L_{patches})$\\
  4. $P$ = $ComputeProbability(encoding)$\\
  5. \uIf{p > 50\%}{
        \textbf{return} "Malicious"
  }
  \Else{
        \textbf{return} "Benign"
  }

\end{algorithm}

Illustrated in Algorithm~\ref{Alg:sherlock} is the overall workflow in our proposed technique to detect/classify an image as a malware. The first step is to resize the image to a size of $224\times224$. Once the standard size image is generated $GeneratePatches$ method takes as input an image of size $224\times224$ and generates linearized patches of size $16\times16$. The list of linearized patches is then provided to our encoder which generates an embedding of the patches $L_{patches}$ capturing complex features. The feature encoding $encoding$ is provided to our pre-trained models which computes the probability $P$ for the encoding which may contain malicious features. If the $P$ is significant (i.e. p > 50\%) the algorithm returns Malware else returns Benign. Similar workflows are followed for type and family inference, with the main difference arising from the number of categories (binary classification for malware detection and multi-class classification for type and family).

We use the transformer based ViT-Base architecture with a patch size of 16 (ViT-B/16). The ViT-Base architecture contains 12 model layers or 'blocks' (as in figure~\ref{transformerencoder}, and 86 million parameters, corresponding to the encoder architecture used for self-supervised learning as described in Section~\ref{sec:MAE_selfsup}. The self-supervised encoder from the self-supervised learning step is reused as-is for each of the 3 downstream classifications (binary, type and family as per Figure~\ref{fig:classifier}). Thus, all the features and weights learned by the self-supervised model are re-used to initialize most of the supervised neural network, which is then further trained to learn improved task-specific features. For each analysis, a linear layer is constructed mapping from the embedding of the encoder (768 in size) to the number of output classes (2, 47 or 696 respectively). The resultant network is fine-tuned end to end (backpropogating gradients and updating weights of every parameter of the neural network), thus enabling the network to learn more specialized features suitable for classifying imagery based on task-specific categories, but building upon the features learned during the self-supervised task of filling in masked patches.

\begin{figure}[h]
\centering
\includegraphics[width=0.5\textwidth]{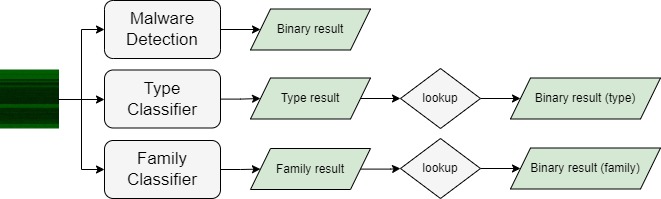}
\caption{Using finer granularity tasks to predict coarse granularity tasks. }
\label{fig:granularity}
\end{figure}

The classes for malware-type and malware-family has a direct one-to-one relationship to the binary class malware-benign. Using this hierarchical relationship we can infer the classification of a given android app as malicious or benign, by training the model to classify for a fine granular class such as malware-type or malware-family. In particular, as shown in Figure~\ref{fig:granularity} we leverage the training from finer granularity classification task to infer a classification for coarse granular class. However, a similar inference cannot be established from malware-family to malware-type in this data-set since there is a many-to-many relationship between these two categories.

\section{Evaluation}
\label{sec:evaluation}
In this section, we evaluate the efficiency of our proposed malware detection method \sherlock ~on the largest open-source dataset for Android malware~\cite{malnetimages}. First, we describe the configuration setup for our experiment, the data set usage and evaluation criteria. Next, we evaluate the performance against state-of-the-art machine learning algorithms, show the confusion matrix of malware classification and discuss the comparison results. 

\begin{table}[t]
\centering
\caption{The number of images and families in each type of malware in MalNet~\cite{malnetimages}}
\label{tab:dataset}
\resizebox{0.4\textwidth}{!}{%
\begin{tabular}{lrrl}
\hline
\textbf{Type} & \multicolumn{1}{l}{\textbf{Img.}} & \multicolumn{1}{l}{\textbf{Fam.}} &  \\
\hline
Adware & 884K & 250 &  \\
Trojan & 179K & 441 &  \\
Benign & 79K & 1 &  \\
Riskware & 32K & 107 &  \\
Addisplay & 17K & 38 &  \\
Spr & 14K & 46 &  \\
Spyware & 7K & 19 &  \\
Exploit & 6K & 13 &  \\
Downloader & 5K & 7 &  \\
Smssend+Trojan & 4K & 25 &  \\
Troj & 3K & 36 &  \\
Smssend & 3K & 12 &  \\
Clicker+Trojan & 3K & 3 &  \\
Adsware & 3K & 16 &  \\
Malware & 3K & 19 &  \\
Adware+Adware & 3K & 2 &  \\
Rog & 2K & 22 &  \\
Spy & 2K & 7 &  \\
Monitor & 1K & 5 &  \\
Ransom+Trojan & 1K & 7 &  \\
Banker+Trojan & 1K & 6 &  \\
Trj & 940 & 18 &  \\
Gray & 922 & 10 &  \\
Adware+Grayware+Virus & 835 & 4 &  \\
Fakeinst+Trojan & 718 & 10 &  \\
Malware+Trj & 609 & 1 &  \\
Backdoor & 602 & 10 &  \\
Dropper+Trojan & 592 & 8 &  \\
Trojandownloader & 568 & 7 &  \\
Hacktool & 542 & 7 &  \\
Fakeapp & 425 & 5 &  \\
Clickfraud+Riskware & 369 & 5 &  \\
Adload & 333 & 4 &  \\
Addisplay+Adware & 294 & 1 &  \\
Adware+Virus & 274 & 9 &  \\
Clicker & 265 & 5 &  \\
Fakeapp+Trojan & 256 & 1 &  \\
Riskware+Smssend & 247 & 7 &  \\
Rootnik+Trojan & 223 & 5 &  \\
Worm & 220 & 7 &  \\
Fakeangry & 211 & 2 &  \\
Virus & 191 & 3 &  \\
Trojandropper & 178 & 4 &  \\
Adwareare & 152 & 3 &  \\
Risktool+Riskware+Virus & 152 & 3 &  \\
Spy+Trojan & 119 & 5 &  \\
Click & 113 & 1 & 
\end{tabular}%
}
\end{table}

\subsection{Experimental Setup}
We evaluate the efficacy of our proposed self-supervised learning method \sherlock~on three tasks using the largest open-source dataset for Android malware~\cite{malnetimages}. First, we evaluate the effectiveness for the pre-training task to synthesise malware images. Second, we evaluate the performance for the classification task of correctly identifying the malware label. Third, we compare our classification performance with existing state of the art tools, and finally we analyze the sensitivity of the synthesis process on the overall classification performance. For comparison with state of the art deep learning architectures we consider; ResNet, DensNet and MobileNetV2 with different configurations. All our experiments were conducted in the Spartan Cluster~\cite{lev2016spartan} on a single node with 24 cores(Intel Xeon CPU E5-2650 v4 @ 2.20GHz) and 4 P100 GPU with 12GB of GPU RAM.

\begin{figure*}[t]
\centering
\includegraphics[width=\textwidth]{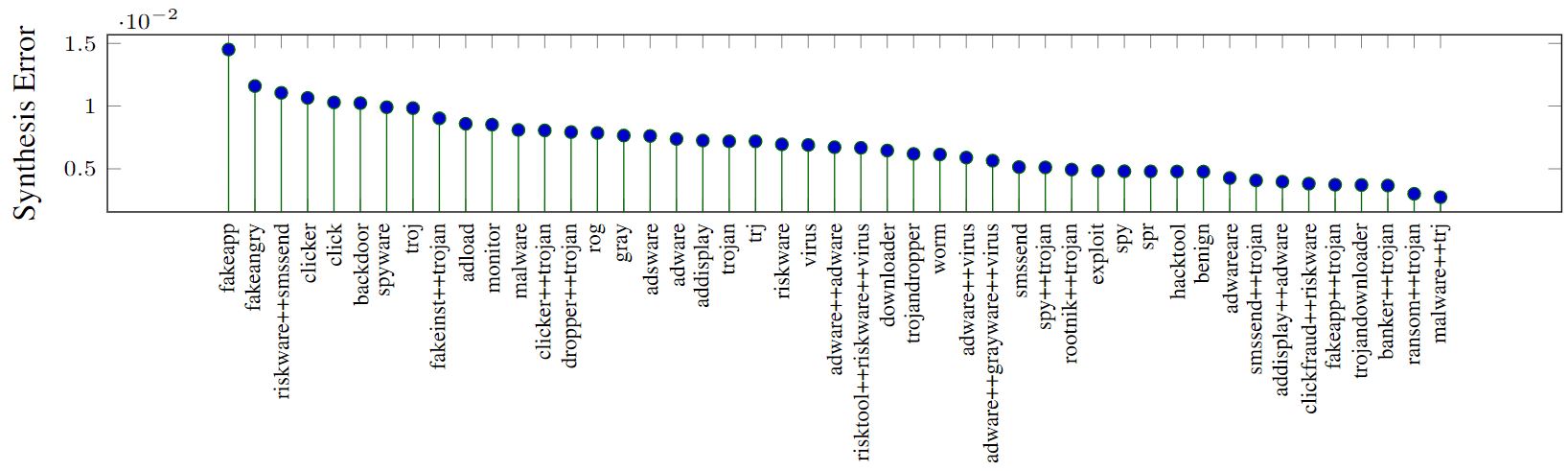}

\caption{Quality of the self-supervised image synthesis process where the reconstruction error is aggregated across malware type}
\label{fig:synthesis-error}
\end{figure*}
For our experiments we use MalNet dataset~\cite{malnetimages} which:
\begin{itemize}
    \item includes 1,262,024 malware images extracted from real-world Android applications in AndroZoo~\cite{androzoo}
    \item contains the largest diverse open-source dataset comprising of 47 malware types and 696 malware families including benign applications
\end{itemize} 

Table~\ref{tab:dataset} provides a detailed breakdown of the number of images and families in each malware type. Notably, the class distribution is highly imbalanced across both image type and family, which is a common property of many real-world datasets, where a few of the classes contain a majority of the examples. For all tasks, we report Accuracy, macro Precision, macro Recall and macro F1 scores, and compare against prior work by utilizing the same train/test splits from this dataset.

\begin{equation}
    Precision = \frac{TP}{TP+FP}
\end{equation}

\begin{equation}
    Recall = \frac{TP}{TP+FN}
\end{equation}

\begin{equation}
    F1 = \frac{2\times Precision \times Recall}{Precision + Recall} 
\end{equation}

Here, TP stands for True Positive, which represents the number of samples belonging to class $c$ which are correctly classified as class $c$; FP stands for False Positive, which represents the number of samples not belonging to class $c$ which are erroneously classified as class $c$; FN stands for False Negative, which is the number of samples that belong to class $c$ which are erroneously classified as a different class. 

For the classifications of malware, type and family, we report the Macro-Precision (MP), Macro-Recall (MR) and Macro-F1 score.

\begin{equation}
\label{eqn:macro-precision}
    Macro-Precision = \frac{\sum_{c}^{}Precision_c}{C}
\end{equation}

\begin{equation}
\label{eqn:macro-recall}
    Macro-Recall = \frac{\sum_{c}^{}Recall_c}{C}
\end{equation}

\begin{equation}
\label{eqn:macro-F1}
    Macro-F1 = \frac{2\times MP \times MR}{MP + MR} 
\end{equation}

The macro criteria are calculated using a class wise average of the metrics as shown in Equations \ref{eqn:macro-precision}, \ref{eqn:macro-recall} and \ref{eqn:macro-F1}. Here $C$ represents the number of classes in each multi-class classification task. Additionally, we will also make use of visualizations to better understand the performance of our models. The confusion matrix as well as the Receiver Operating Characteristic (ROC) curve will be used to understand the nuances of our model performance.

\begin{equation}
\label{eqn:rebalance}
    W_c = \frac{1-0.999}{1-0.999^{n_c}}
\end{equation}

For the training process, we used simple cross entropy loss with class re-balancing. The classes which have fewer training examples in the dataset are weighted higher in the loss function. We use the weights suggested in \cite{malnetimages} as shown in Equation~\ref{eqn:rebalance}, where $n_c$ is the number of images in class $c$. These weights cause a relatively higher loss to be assigned to miscategorizations for samples originating from under-represented classes. Therefore, the neural network will prioritize learning a representation that can more accurately classify such under-represented classes, compared to a standard (unweighted) cross-entropy loss.



\subsection{Accuracy of Malware Image Synthesis}

We quantify the ability of the self-supervised representation to regenerate the images in the test set. As this is a direct evaluation of the self-supervisory task, it can be interpreted as an initial evaluation of the ability of the representation and decoder to capture the semantic features associated with the binaries of the considered categories.

For each class in malware-type we iterate over each image and generate 10 random masks for 75\% of the image and generate 10 masked input images containing only 25\% of the pixels from the original image. For each masked input image, using the trained self-supervised model we generate an image that is similar to the original image provided. For each synthesised image we compute the absolute error for each pixel in comparison to the original image, and average the error across all 10 randomly generated masked input images. Finally, the error is averaged across all such images for each class. The aggregated reconstruction error across malware type is shown in Figure~\ref{fig:synthesis-error}. 


\begin{table*}[t]
\centering
\caption{Results on malware classification on 3 classes against 3 popular architectures ResNet, DenseNet and MobileNetV2—on macro-F1, macro-precision, and macro-recall. Results for compared architectures are referenced from MalNet dataset~\cite{malnetimages}}
\label{table:results-classification}
\resizebox{\textwidth}{!}{%
\begin{tabular}{|l|ccc|ccc|ccc|}
\hline
\multicolumn{1}{|c|}{\multirow{2}{*}{\textbf{Model}}} & \multicolumn{3}{c|}{\textbf{Binary}}                                                         & \multicolumn{3}{c|}{\textbf{Type}}                                                           & \multicolumn{3}{c|}{\textbf{Family}}                                                         \\ \cline{2-10} 
\multicolumn{1}{|c|}{}                                & \multicolumn{1}{c|}{F1 Score} & \multicolumn{1}{c|}{Precision} & \multicolumn{1}{c|}{Recall} & \multicolumn{1}{c|}{F1 Score} & \multicolumn{1}{c|}{Precision} & \multicolumn{1}{c|}{Recall} & \multicolumn{1}{c|}{F1 Score} & \multicolumn{1}{c|}{Precision} & \multicolumn{1}{c|}{Recall} \\ \hline
ViT-B (SHERLOCK)                                    & .854                                  & \textbf{.920}                          & .810                        & \textbf{.497}                           & \textbf{.628}                          & \textbf{.447}                         & \textbf{.491  }                         & \textbf{.568}                              & \textbf{.461}                                \\
ResNet18                                            & .862                                  & .893                                   & .837                                & .467                                  & .556                                   & .424                                & .454                                  & .538                                   & .423                                \\
ResNet50                                            & .854                                  & .907                                   & .814                                & .479                                  & .566                                   & .441                                & .468                                  & .541                                   & .443                                \\
DenseNet121                                         & .864                                  & .900                                   & .834                                & .471                                  & .558                                   & .428                                & .461                                  & .529                                   & .438                                \\
Densenet169                                         & \textbf{.864}                                  & .890                          & \textbf{ .841}                                & .477                                  & .573                                   & .433                                & .462                                  & .545                                   & .434                                \\
MobileNetV2(x.5)                                    & .857                                  & .894                                   & .827                                & .460                                  & .547                                   & .424                                & .451                                  & .528                                   & .423                                \\
MobileNetV2(x1)                                     & .854                                  & .889                                   & .825                                & .452                                  & .527                                   & .419                                & .438                                  & .532                                   & .405   \\
\hline
\end{tabular}%
}
\end{table*}

The reconstruction error aggregated over malware-type ranges from 0.27\%-1.45\% with an overall average of 0.68\%. It is useful to note that the scale of the error computed depends on the scale of pixel values in the image (ranging between [0, 255] or [0.0, 1.0] per channel, as we use in this work) and the proportion of masked pixels in the image. The same averaging procedure is used across categories and the masked pixel percentage is constant for each image, the final averages we generate are comparable across categories (and only differ from other similar evaluations by some scalar factor). However, if the masked percentage were to differ between images, it would be necessary to take the absolute error between only the masked pixels of the original image and the synthesized image.

\begin{tcolorbox}[boxrule=1pt,left=1pt,right=1pt,top=1pt,bottom=1pt]
\textbf{Malware Image Synthesis Accuracy:}
ViT-Base architecture we used in \sherlock~was able to accurately synthesise malware images of 75\% masking with an average reconstruction error of 0.68\%.
\end{tcolorbox}

\subsection{Malware Classification}
We evaluate the efficacy of our proposed self-supervised model \sherlock~in correctly identifying the label for each malware-image in 3 different categories. For each category we train a separate classifier for which semantic learning from the supervised-learning is transferred as shown in Figure~\ref{fig:classifier}. For each category we report the macro-F1, macro-Precision and macro-Recall as shown in Table~\ref{table:results-classification}. For comparison we use 3 popular deep learning architectures--ResNet, DenseNet and MobileNetV2. The results for the compared architectures are referenced from the MalNet dataset~\cite{malnetimages}. 

\sherlock~was able to correctly classify the malware-type and malware-family across 47 and 696 classes respectively. Our model was able to classify both classes correctly with an accuracy of 83.7\% for malware-type and 80.2\% for malware-family, outperforming the state-of-the-art deep learning models. Furthermore, \sherlock~recorded the highest macro-F1 score, macro-precision and macro-recall in both categories indicating the superiority of using a self-supervised learning model over existing supervised learning models. One of the contrasting differences between the compared models and \sherlock~is the source for transfer learning. \sherlock~is able to directly learn the semantics of a malware image due to its self-supervised pre-training task while the compared models are trained from scratch. Prior work~\cite{malnetimages} has used transfer learning from ImageNet~\cite{deng2009imagenet} to pretrain a model and then fine tune to the training data, however it performs significantly worse than the one trained from scratch due to the difference of the semantics between ImageNet images and MalNet images.

\begin{figure*}[t]
\includegraphics[width=1\textwidth,clip,keepaspectratio]{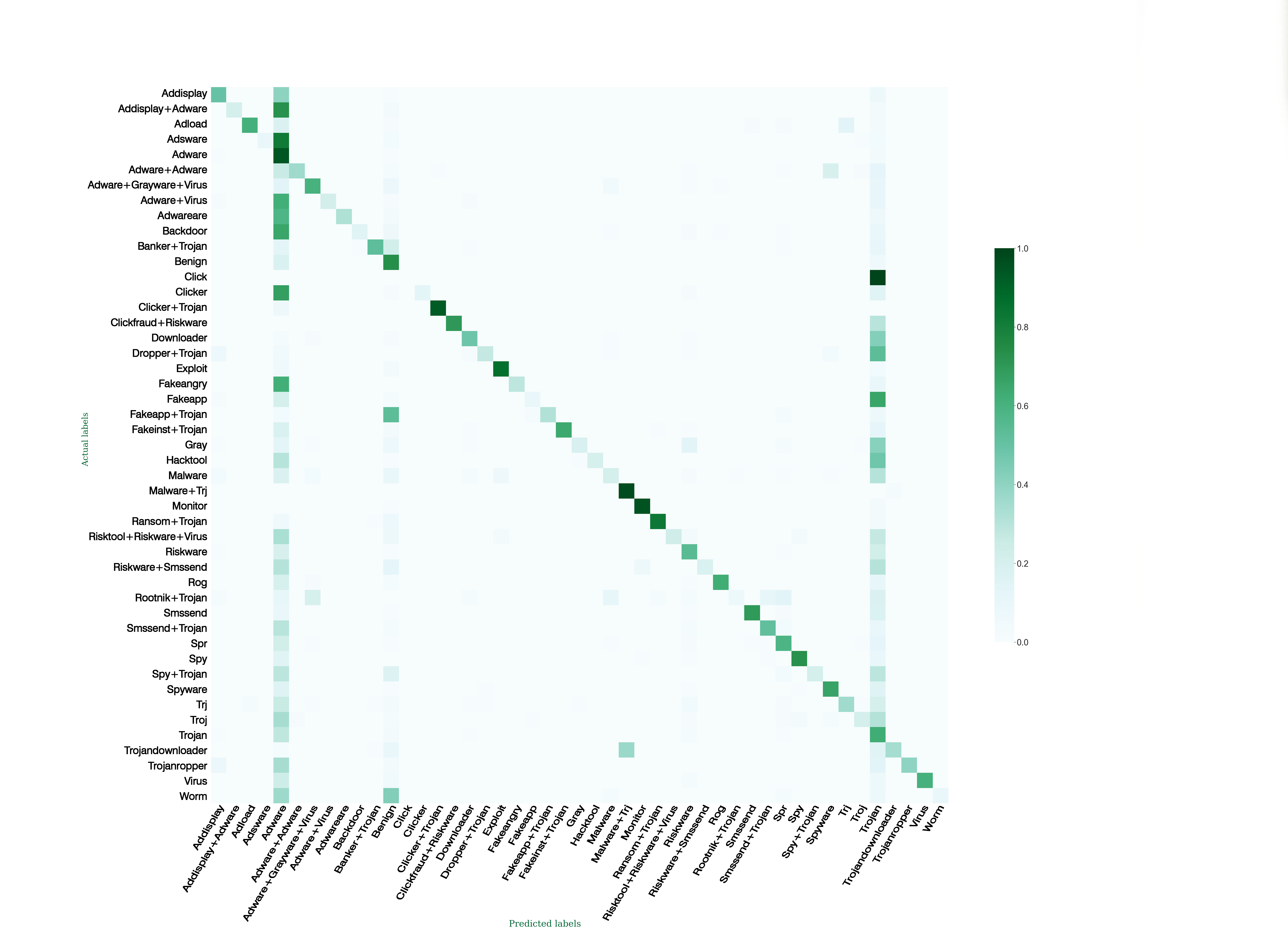}
\caption{Confusion Matrix for malware type classification, which visualize the performance in correctly identifying the actual class label. The darker the shade the stronger the classifier performance and lighter-shade indicates poor performance. Our model results depict more darker shades for the diagonal entries while lighter-shade for non-diagonal entries indicating good performance.}
\label{figure:matrix-binary}
\end{figure*}


Despite the performance for malware-type and malware-family, our model \sherlock~does not perform well for the binary classification task. Our model reports the best performance in precision while the worse performance in recall for the binary classification task, resulting in worse overall performance in macro-F1 score. Cross comparison with the performance of our model in all 3 tasks, indicates the precision of the results are improved from the semantics learnt from the self-supervised learning, while recall can be improved with additional fine-tuning. However since the class imbalance is high (i.e. 2:31) for the binary classification task the fine-tuning does not have access to enough training data to accurately learn features for the benign class. However, with more classes introduced in the tasks for classifying malware-type and malware-family the imbalance effect is distributed over multiple classes, and improves overall recall. This indicates a strong influence from the pre-training task which learns semantic information for the secondary classification task. Since the pre-training task in our model is optimized to synthesise images it learns necessary semantic features relevant for the synthesis. Additionally due to the class imbalance problem in our dataset the semantic learning is skewed towards malware images, resulting in a higher number of misclassifications as malware.


\begin{tcolorbox}[boxrule=1pt,left=1pt,right=1pt,top=1pt,bottom=1pt]
\textbf{Malware Classification Accuracy:}
\sherlock~was able to outperform existing state-of-the-art deep learning models for malware-type and malware-family with an macro F1-Score of .497 and .491 respectively. 
\end{tcolorbox}


\begin{table}[h]
\centering
\caption{Efficacy of malware detection against 3 popular architectures ResNet, DenseNet and MobileNetV2—on its accuracy, macro-F1, macro-precision, and macro-recall.}

\resizebox{0.48\textwidth}{!}{%
\begin{tabular}{lrrccc}
\hline
\textbf{Model}  & \textbf{F1 Score} & \textbf{Precision} & \textbf{Recall} \\
\hline

\sherlock & .854 & \textbf{.920} & .810  \\
\sherlock (type) & .876 & .891 & .862  \\
\sherlock (family)  & \textbf{.878} & .867 & \textbf{.890}  \\
ResNet18 &  .862 & .893 & .837\\
ResNet50 &  .854 & .907 & .814\\
DensetNet121 & .864 & .900 &.834 \\
DenseNet169 &  .864 & .890 & .841\\ 
MobileNetV2(x.5) &  .857 & .894 & .827\\
MobileNetV2(x1) & .854 & .889 & .825\\
\hline

\end{tabular}%
}
\label{table:results-detection}
\end{table}

\subsection{Malware Detection}
We evaluate the overall effectiveness of \sherlock~to accurately detect a malware application using inference based on the classification label it generates as described in Figure~\ref{fig:granularity}. Importantly, generating an image from the corresponding binary takes on average 0.479 seconds, while inferring a label using Sherlock takes on average 0.003 seconds, for a total inference time of 0.482 seconds with a single 4 GPU node containing 24 cores(Intel Xeon CPU E5-2650 v4 @ 2.20GHz) and 4 P100 GPU with 12GB of GPU RAM each on Spartan\cite{lev2016spartan}. A comparative analysis of our inference result with the state-of-the-art deep learning models is presented in Table~\ref{table:results-detection}, where the results for state-of-the-art is extracted from recent work~\cite{malnetimages}. Our model \sherlock~has the best performance in all three dimensions in macro F1-score, macro precision and macro recall. However, inference using malware-family classifier has the overall best-performance with an macro F1-score of 0.878, indicating that pre-trained model with finer granularity can learn distinct features to differentiate an image between the benign class and malware class.

\begin{figure}[h]
\begin{tikzpicture}
 
\begin{axis} [xbar = .05cm,
    bar width = 0.3cm,
    width= 0.48\textwidth,
    major grid style={draw=white},
    font = \tiny,
    height= 5cm,
    xmin = 80,
    xmax = 95,
    ytick = data,
    enlarge y limits=0.2,
    nodes near coords,
    xlabel={Percentage(\%)},
    symbolic y coords={Recall,Precision,F1},
    legend cell align = {right},
    legend style={
            at={(0.5,-0.3)},
            font=\tiny,
            anchor=north,
            legend columns = 3,
    },
    legend entries = {SHERLOCK,SHERLOCK(type),SHERLOCK(family)}
]
 
\addplot [draw=none, fill=blue!70] coordinates {(85.4,F1) (92,Precision) (81,Recall)};
\addplot [draw=none,fill=black!30!green]  coordinates {(87.6,F1) (89.1,Precision) (86.2,Recall)};

\addplot  [draw=none, fill=black!10!orange]  coordinates {(87.8,F1) (86.7,Precision) (89,Recall)};
 
 
\end{axis}
 
\end{tikzpicture}
\caption{In-depth analysis of the performance metrics of \sherlock~with 3 binary classification models trained on different granularity}
\label{fig:detection-sherlock-only}
\end{figure}
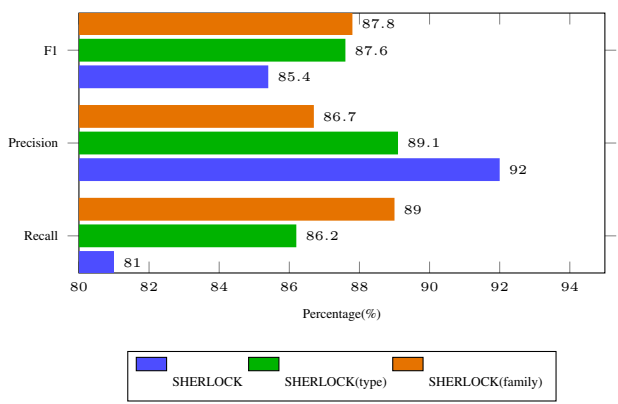

\begin{figure}[h]
\centering
\includegraphics[width=0.5\textwidth]{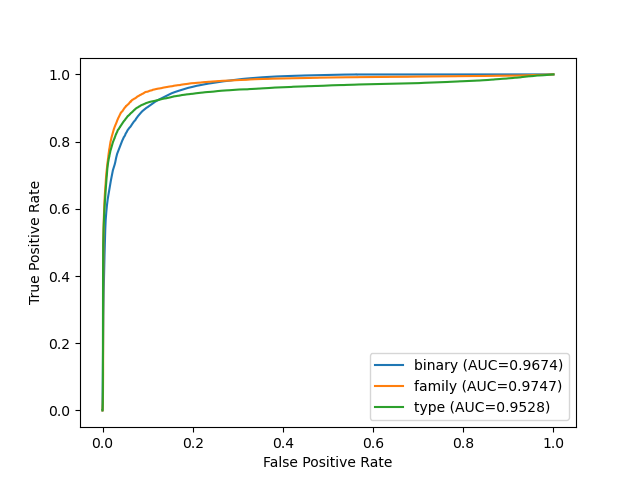}
\caption{Receiver Operating Characteristics (ROC) curve for malware detection with 3 binary classification models trained on different granularity }
\label{fig:roc}
\end{figure}

For an in-depth analysis we plot the results of our inference models in Figure~\ref{fig:detection-sherlock-only}. Binary classification of a given byte-plot image trained with coarse-granularity achieves highest precision but lowest recall. For applications such as Malware detection where precision is most important, a model pre-trained with self-supervised learning can achieve state-of-the-art performance, with the final model trained on coarse-grained task such as two-class labeling. Whereas the same pre-trained model with the final model trained on a finer-grained task such as identifying the malware-type/malware-family (N classes) will be more effective in improving recall and overall F1-score. An observation we make is that with the increasing number of classes for the final model, the recall is improved while precision is decreased. Additionally, both malware-type based inference and malware-family based inference is able to achieve similar macro F1-score with the contrasting differences in its precision and recall values. While malware-type based inference has the better precision in comparison with malware-family, recall is better for the latter. To further understand the significance of our results, we also plot the Receiver Operating Characteristics (ROC) curve in Figure~\ref{fig:roc}. Despite having a better F1-score, malware-type based inference has a low Area Under Curve (AUC), indicating the classifier will not perform well for different thresholds. However, malware-family based inference model has the highest AUC score as well as for the macro F1-score.

\begin{tcolorbox}[boxrule=1pt,left=1pt,right=1pt,top=1pt,bottom=1pt]
\textbf{Malware Detection Efficacy:}
\sherlock~was able to correctly detect a malware with an accuracy of 97\% for an highly imbalanced dataset while maintaining 86.7\% precision and 89\% recall. 
\end{tcolorbox}

\newpage
\section{Related Work}
\label{sec:related}

\subsection{Malware Detection and Classification}
\begin{figure*}[t]
\centering
	{\small \begin{forest}
			for tree={
				align=center,
				parent anchor=south,
				child anchor=north,
				edge path={
					\noexpand\path [draw, \forestoption{edge}] (!u.parent anchor) -- +(0,-10pt) -| (.child anchor)\forestoption{edge label};
				},
				if level=0{
					inner xsep=0pt,
					tikz={\draw [thick] (.south east) -- (.south west);}
				}{}
			}
			[Malware Detection and Classification
			[Features
			[Static]
			[Dynamic]
			[Hybrid]]
			[Algorithms
			[Signature Based]
			[Artificial Intelligence approaches
			[Supervised]
			[Unsupervised]
			[Semi-supervised]
			]
			]
			]
	\end{forest}}
	\caption{Categorization of malware detection and classification \cite{abusitta2021malware}}
	\label{fig:cat}
\end{figure*}
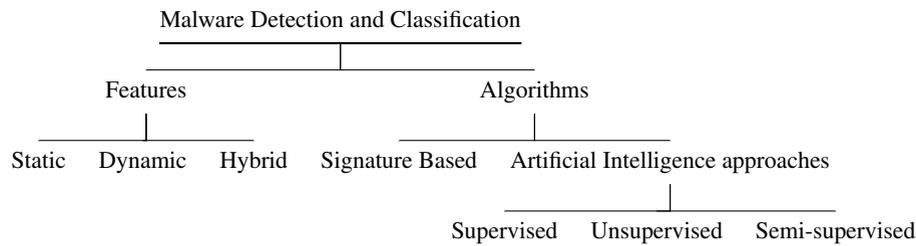

Malware detection and classification has become a crucial task due to the increasing complexity of malware and the commonality of computing systems. Abusitta et. al. \cite{abusitta2021malware} categorized malware detection and classification approaches based on features and algorithms used, as shown in Figure~\ref{fig:cat}.

Features used for the analysis can be classified as static or dynamic. Static features are extracted from the executable files while dynamic feature extraction methods make use of a sandbox environment to run the program and collect the memory image or the behaviors of the program execution to extract features \cite{damodaran2017comparison}. Some of the features in literature include printable strings \cite{huang2016mtnet, dahl2013large, lee2020cross}, byte code \cite{anderson2012improving, saxe2015deep, amin2020static, daoudi2021dexray}, assembly code \cite{anderson2011graph, dai2009efficient, shankarapani2011malware}, API/DLL system calls \cite{shankarapani2011malware, sami2010malware, salehi2014using}, control flow graphs \cite{ma2019combination, yan2019classifying, bruschi2006detecting} and function level features \cite{cai2021learning}. Algorithm used for detection and classification approaches can be broadly categorized into signature based approaches and Artificial Intelligence based approaches. Signature based approaches are the de facto of current antivirus providers \cite{al2019review}. These signatures are created by humans and the malware detectors performs a matching between programs and signatures to detect or classify malware \cite{schultz2000data, christodorescu2003static}. One can argue this method could only detect known malware and dependent on the manual signature generation.  There is a tendency to move towards AI based malware detection and classification which can mainly be categorized into supervised \cite{singh2022assessment, ki2015novel}, unsupervised \cite{tang2014unsupervised, liu2021research, fan2019graph} and semi supervised \cite{santos2011semi, mahindru2020feature, gao2020malware} methods. In supervised malware detection the model learns features from a labeled dataset, while unsupervised methods extract patterns from unlabeled data. Semi-supervised methods make use of both labeled and unlabeled data. While traditional methods like Naïve Bayes, Decision Trees, K-Nearest Neighbors and SVM have been extensively used in the past. An analysis of these classical methods showed SVM performed well for malware detection \cite{souri2018state}. An android specific malware detection survey by Liu et. al looks at the current ML methods used for malware detection \cite{liu2020review}. More recently ReLU, LSTM and CNN based methods have received more attention. A deep learning based detection method which works with image data was explored by Yadav et. al. \cite{yadav2022two}. Due to the ability to recognize previously unseen malware and the higher performance over traditional methods, these machine learning based malware detection techniques are becoming widely researched \cite{ahvanooey2020survey}.

Our approach can be classified as a self-supervised learning based algorithm that make use of static features of the application. Contrast to existing work, this is the first self-supervised based learning method that has been studied with the largest openly-available dataset for Android malware.

\subsection{Image Classification}
Imagery such as those generated to represent code or program binaries broadly fall into the category known as \textbf{synthetic imagery}. Such imagery is distinct from \textbf{natural imagery}, which is the standard type of data used in computer vision in datasets such as ImageNet\cite{deng2009imagenet}, incorporating images of real world scenes captured through a camera. Exploration of the performance of self-supervised methods on complex synthetic imagery remains under-explored compared to natural imagery. Some prior work in this area has explored representation learning from  sketches\cite{xie2019unsupervised,bhunia2021vectorization}, altitude imagery\cite{seneviratne2021contrastive}, synthetic scene imagery\cite{ren2018cross}, and Google map imagery\cite{seneviratne2021self}. Analysis in such domains contributes useful knowledge regarding the performance of self-supervised methods on synthetic imagery, which have different visual features to natural imagery.

\section{Perspectives}
\label{sec:perspectives}

A key advantage of using self-supervision in this imagery domain (malware imagery) is the ability to initialize a single general representation in a task-independent manner. This representation can then be specialized to perform specific analysis (such as malware detection, malware type categorization and malware family categorization, as well as potentially other tasks). In essence, the representation can be easily specialized for any different set of labels associated with the imagery, for example future work could look at labels associated with malware authors and attempt to identify unique "signatures" corresponding to authors from the imagery.

Interestingly, the use of coarser labels has improved performance in the malware detection task (classification of malware images into benign vs malicious). Both malware family and type predictions can be converted into a malware detection prediction by performing a lookup. We found that the models for family and type prediction, once converted into the corresponding malware detection category ("malicious" or "benign") provided superior performance when compared with the malware detection prediction, when all models were trained with the same number of images and identical analytical settings. Since the only difference between the analyses was the granularity of the labels (696 for family, 47 for type, 2 for malware detection), we can conclude that the coarser labels have led to a better characterization of the images for the malware detection task. In this aspect, while both models trained on more granular labels show significant improvement in F1 score over the malware detection model, they both have similar performance to each other (in terms of F1 score). This suggests that the improvement gained via increased label granularity has diminishing returns, which may be interesting for future work to explore.

Upon further consideration, the reason for this improvement can be induced. On a fixed dataset, with fine-grained labels, the model is forced to create multiple decision boundaries to separate instances from different categories from each other. The number of potential required boundaries will grow quadratically in the number of categories (n*(n-1)/2), as each category may be adjacent to every other in the feature space. This leads to a more granular clustering of images with a larger number of clusters containing fewer images in the feature space (as the total number of images is fixed in all analyses). 
Additionally, this provides further reasoning for the muted improvement (in terms of malware detection results) of increasing the number of categories from 47 to 696. The malware detection task has 1 decision boundary (benign vs malicious), whereas type prediction 47*46/2 = 1081 potential boundaries and family prediction has 676*675/2 = 228150. Based on the observed results the granularity of type labels is sufficient to capture the level of detail required for the binary classification required in malware detection, and the added granularity of the family labels does not improve results further. The impact of the label granularity is also apparent in the ROC curves for malware detection, with the more granularly labelled family prediction model showing superior performance to the other more coarsely labelled models. Future work could explore this aspect further by exploring the added impact of having a more complex baseline task (with more than 2 categories). As the type and family labels do not have a one to one mapping, we are unable to explore this any further at this stage. 
The impact of label granularity of classification has previously been studied in convolutional neural networks\cite{chen2018understanding}. The findings in our work further generalize such results to the Vision Transformer architecture.

\section{Conclusion}
\label{sec:conlusion}
In this paper, we present a system based on self-supervised learning for malware classification, \sherlock which is a transformer based computer vision model that utilizes self-supervised learning to detect Android malware. We evaluate \sherlock ~on a large-scale data-set of 1.2 million Android apps consisting of 47 malware types and 696 malware families. Compared with state-of-the-art deep learning architectures ResNet, DenseNet and MobileNetV2, our proposed technique \sherlock ~outperforms all techniques in malware detection and classification with 97\% accuracy for detection. \sherlock ~was also able to correctly classify malware type and malware family with a macro-F1 score of \macrotype~ and \macrofamily~ respectively. This work demonstrates the efficacy of self-supervised computer vision models, specifically the Vision Transformer architectures in the use for malware classification. Through this work, we improve static analysis based malware detection and classification, which we believe can be further improved with by augmenting additional features derived from dynamic analysis. 

To allow the research community to better replicate and reproduce the findings in our studies and to extend the work further, we have open-sourced our model to the community in Github: \href{https://github.com/sachith500/Sherlock}{https://github.com/sachith500/Sherlock}.

\section*{Acknowledgments}

This project was supported by Australian NHMRC Grant GA80134. This research was undertaken using the LIEF HPC-GPGPU Facility hosted at the University of Melbourne. This Facility was established with the assistance of LIEF Grant LE170100200.

\bibliographystyle{IEEEtran}
\bibliography{references}

\begin{IEEEbiography}[{\includegraphics[width=1in,height=1.25in,clip,keepaspectratio]{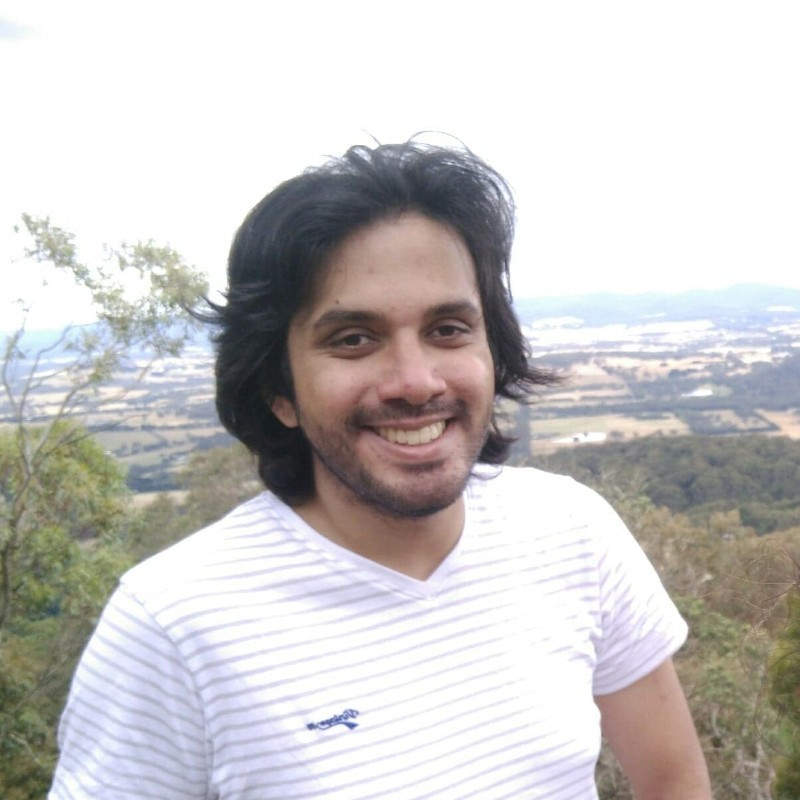}}]{Sachith Seneviratne}
received his B.Sc. degree in computer science and engineering from the University of Moratuwa, Sri Lanka. He completed his PhD in machine learning from Monash University, Australia. Currently, he is working as a research fellow at the University of Melbourne. Sachith's research work revolves around deep learning, with a focus on contrastive representation learning and applications. He is broadly interested in self-supervised deep learning approaches across various disciplines such as computer vision, NLP and reinforcement learning.

\end{IEEEbiography}

\begin{IEEEbiography}[{\includegraphics[width=1in,height=1.25in,clip,keepaspectratio]{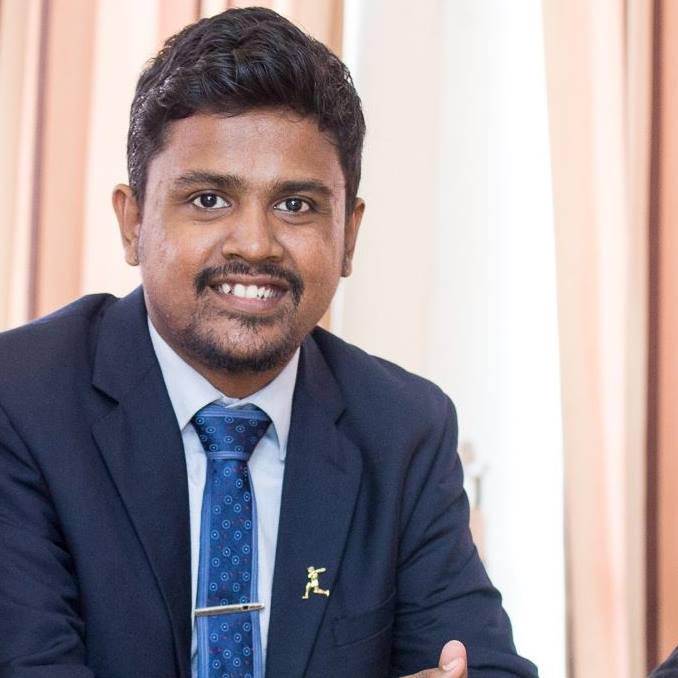}}]{Ridwan Shariffdeen} is a PhD Candidate in Department of Computer Science at School of Computing, National University of Singapore. His current focus of research is on automated program repair, software security and software engineering automation.  He received his Bachelor of Science(BSc) Honors (Hons) degree from the Computer Science \& Engineering Department at University of Moratuwa, Sri Lanka. From 2016-2018 he worked in the industry as a Senior Software Engineer for System Security and Automation Technology. 
Mr. Ridwan's research has been recognized by the National Research Foundation Singapore with an invitation to the 10th Global Young Scientist Summit (GYSS'22). He was also awarded the Research Achievement Award by School of Computing, National University of Singapore in 2021 and CINTEC Award for the best Computer Science and Engineering graduand for Research by University of Moratuwa, Sri Lanka in 2016.

\end{IEEEbiography}

\begin{IEEEbiography}[{\includegraphics[width=1in,height=1.25in,clip,keepaspectratio]{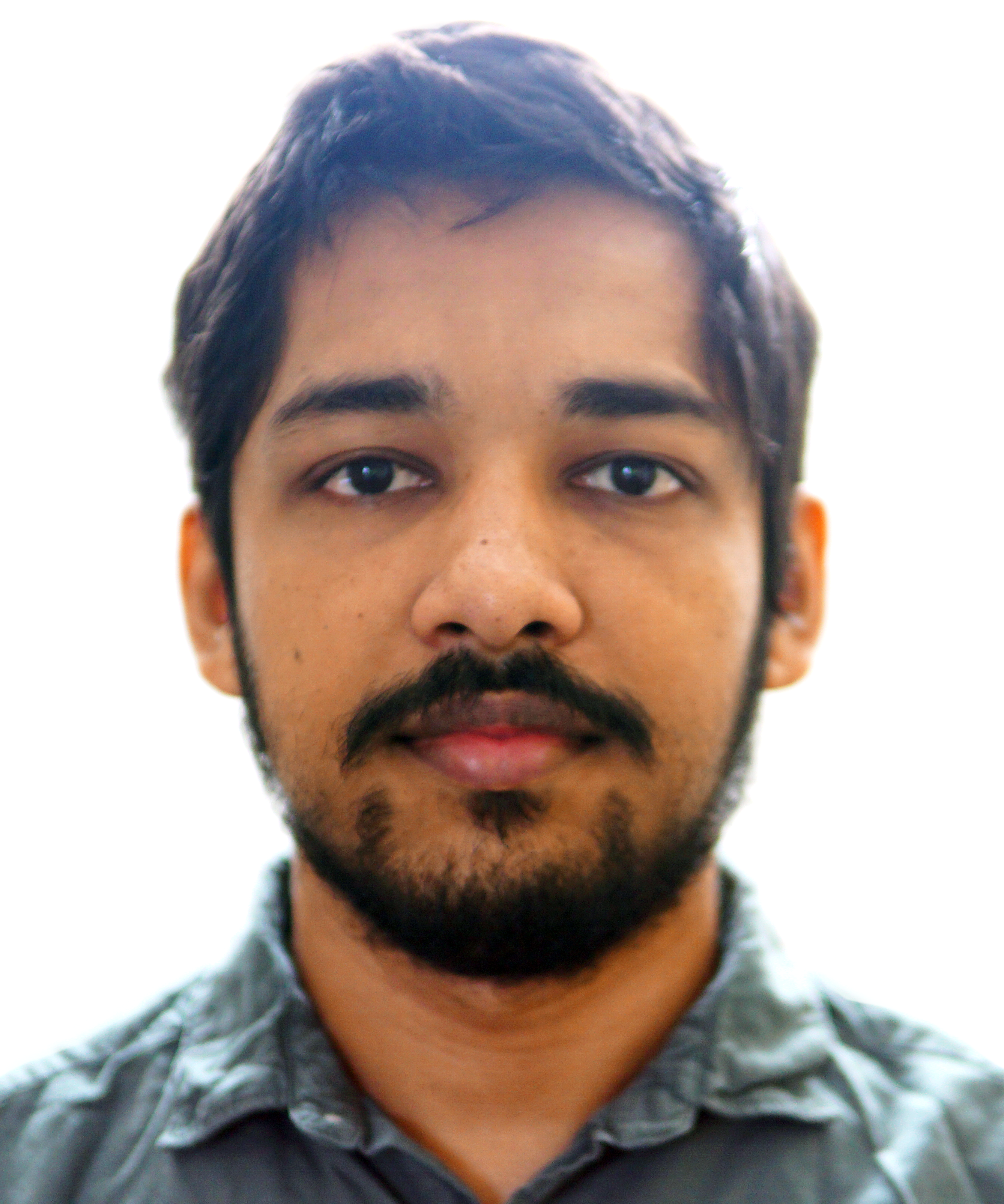}}]{Sanka Rasnayaka} is a lecturer at the School of Computing in the National University of Singapore. His research interests lie in applications of AI and computer vision. He mainly works in the fields of Biometrics and authentication.
In 2021, he received his PhD in Computer Science from the School of Computing in the National University of Singapore, for his work on Continuous Authentication for mobile devices. Prior to that in 2016 he received his Bachelor of Science Honors degree in Computer Science and Engineering from the Engineering Faculty of the University of Moratuwa, Sri Lanka.

\end{IEEEbiography}

\begin{IEEEbiography}[{\includegraphics[width=1in,height=1.25in,clip,keepaspectratio]{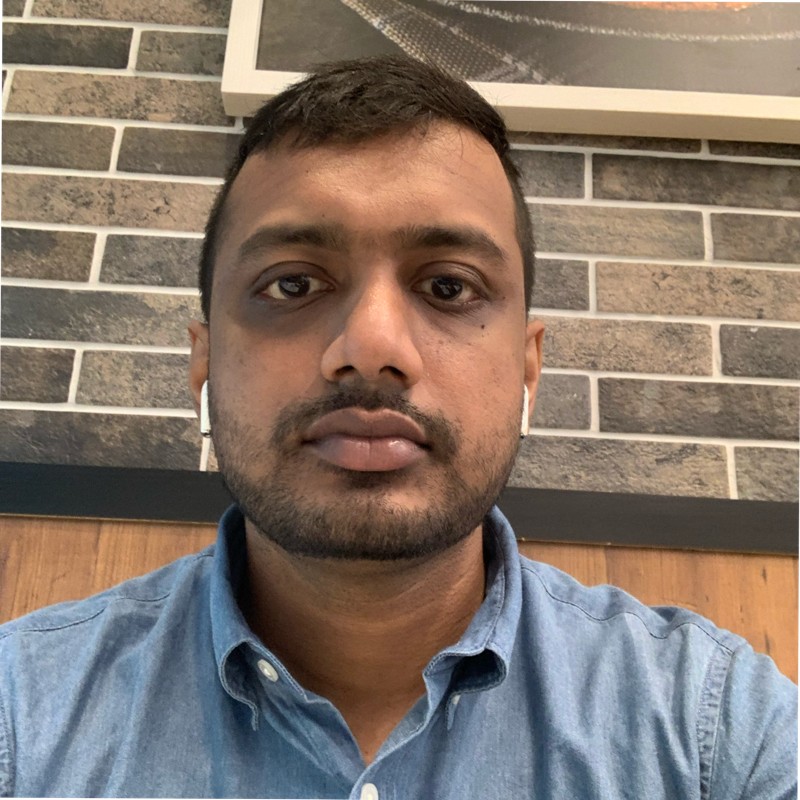}}]{Nuran Kasthuriarachchi} is an Artificial Intelligence Engineer at the Singapore Press Holdings. He has multiple years of experience in the industry working in the fields of Aritifical Intelligence and Machine Learning. He obtained his Bachelor of Science (BSc) Honors degree from the Computer Science and Engineering department of the University of Moratuwa, Sri Lanka in 2016.
\end{IEEEbiography}

\EOD

\end{document}